\documentclass[12pt,draftclsnofoot, onecolumn]{IEEEtran}
\usepackage{amsmath, amsthm, amssymb}
\usepackage{amsthm}

\usepackage{subfig}
\usepackage{cite}
 \usepackage[dvips]{graphicx}

\newtheoremstyle{note}% name
  {3pt}%      Space above
  {3pt}%      Space below
  {\itshape}%         Body font
  {}%         Indent amount (empty = no indent, \parindent = para indent)
  {\itshape}% Thm head font
  {:}%        Punctuation after thm head
  {.5em}%     Space after thm head: " " = normal interword space;
  {}%         Thm head spec (can be left empty, meaning `normal')

\theoremstyle{note}

\newtheorem{theorem}{Theorem}
\newtheorem{corollary}{Corollary}

\begin{document}

\title{Performance of Optimum Combining in a Poisson Field of Interferers and Rayleigh Fading Channels}

\author{Olfa~Ben Sik Ali, %~\IEEEmembership{Member,~IEEE,}
        Christian~Cardinal,~\IEEEmembership{Member,~IEEE,}
        and~Fran\c{c}ois~Gagnon,~\IEEEmembership{Senior~Member,~IEEE}% <-this % stops a space
\thanks{O. Ben Sik Ali and C. Cardinal are with the Department
of Electrical Engineering, \'Ecole Polytechnique, Montr\'eal,
CA,  (e-mail: \{olfa.ben-sik-ali, christian.cardinal\}@polymtl.ca).}% <-this % stops a space
\thanks{F. Gagnon is with the Department
of Electrical Engineering, \'Ecole de Technologie Sup\'erieure,
Montr\'eal, CA,  (e-mail: Francois.Gagnon@etsmtl.ca ).}}% <-this % stops a space

%\thanks{Manuscript received April 19, 2005; revised January 11, 2007.}}

\markboth{Submitted To  IEEE Transactions On Wireless Communications} %Journal of \LaTeX\ Class Files,~Vol.~6, No.~1, January~2007}%
{Shell \MakeLowercase{\textit{et al.}}: Bare Demo of IEEEtran.cls for Journals}

\maketitle

\begin{abstract}
This paper studies the performance of antenna array processing
 in distributed multiple access networks without power control. The
 interference is represented as a Poisson point process. Desired and interfering
signals are subject to both path-loss fading (with an exponent
greater than $2$) and to independent Rayleigh fading. Using these
assumptions, we derive the exact closed form expression for the
cumulative distribution function of the output
signal-to-interference-plus-noise ratio when optimum combining is
applied. This results in a pertinent measure of the network
performance in terms of  the outage probability, which in turn
provides insights into the network capacity gain that could be
achieved with antenna array processing. We present and discuss
examples of applications, as well as some numerical results.

\end{abstract}


\IEEEpeerreviewmaketitle

\section{Introduction}

 \IEEEPARstart{I}n wireless communications, energy and spectrum resources are scarce
and expensive, and must consequently  be managed efficiently  in
order to address the growing requirements of emerging
applications. Intensive work has been dedicated to developing
advanced processing technologies to improve the spectral
efficiency. In decentralized multiple access systems, such as ad
hoc networks, interference is the major performance inhibitor,
which explains why in recent years, much research effort has been
dedicated to the  interference mitigation. Several solutions have
been proposed involving  exploitation of  the particular structure
of the interference. In fact, technologies such as spread spectrum
and multiple antennas can be used to design systems with a number
of degrees of freedom that, when properly exploited, increase
system tolerance to interference. In the context of spread
spectrum, various multiuser receiver schemes have been introduced
\cite{verdu}. The well-known minimum mean square error (MMSE)
receiver  is the linear multiuser receiver that maximizes the
signal-to-interference-plus-noise ratio (SINR), and the equivalent
solution in the context of antenna array processing is known as
the optimum combining (OC) receiver \cite{bair}. Though the study
of the reliability of these receivers is an important issue, which
has been the subject of extensive work \cite{ad1,ad2,ad3,ad4},
there is still a lot of work left to  do in order to understand
their limits in various contexts. In fact, the bulk of  the
results available deal with special assumptions concerning
interference, such as considering equal power interferers, strong
interferers or asymptotic cases (infinite number of interferers
and antennas). In this paper, we address the issue of quantifying
the performance gain that could be achieved by employing the
optimum combining receiver in a decentralized network without
using power or access control.

\subsection{Problem statement and contribution}

As previously mentioned, we are interested in distributed random
access networks. A common representation of this kind of  networks
 is to consider a random number of interferers that
are independently and identically distributed over the area. The
number of interferers is measured by one parameter, namely, their
spatial density. Thus, we adopt the so-called stochastic geometric
model  which is widely used in the literature for the access layer
as well as for the physical layer \cite{sous,weber,Baccelli}. In
this model, the link outage
  is  defined as the probability that the expected
SINR  seen by a representative receiver is below a certain
threshold. The expectation is  taken over the set of possible
realizations of the network. As defined, the outage probability
serves to derive several spatial performance metrics, namely, the
mean number of transmissions in outage per unit area,  the mean
throughput per unit area, and the mean distance covered by all
transmissions per unit area. From these spatial performance
metrics, we can study the connectivity, the transport capacity and
several important issues of the network performance.  The tutorial
paper \cite{hag} provides a survey of these issues and discusses
some applications of this model and this methodology to wireless
communication.\\
To derive the outage probability, we need   the interference
distribution. In a stochastic geometric network, the interference
is a spatial shot-noise process with an impulse response having a
decaying  power  law form. Even with one antenna and the matched
filter receiver, the distribution of the interference does not
take a closed form (except for some special cases, such as a path
loss exponent equal to $4$) although its characteristic function
is well known
\cite{rice,wil,lowen,syn}. \\
Recent works have focused on analyzing the performance  of more
advanced receivers  than the conventional one. In
\cite{weber,olfa} and \cite{serkey}, approximations and  bounds on
the outage probability are given for the successive interference
cancellation receiver. In the context of smart antennas,
\cite{weber3} gives lower and upper bounds on the outage
probability for sectorized antenna, maximal ratio combining and
space time coding techniques. Analyses of the zero-forcing and the
partial zero-forcing receivers follow in \cite{weber4,weber6}.
 The main contribution of the work presented in this paper  is the exact derivation of the outage probability for the optimum combining receiver with
an arbitrary number of antennas.

\subsection{Organization of the paper}
 The outline for this paper is as follows. In
Section II, we present the system model. The derivation of the
outage probability, which is the main result obtained in  this
paper, is presented in Section III. Applications and simulations
follow in Sections IV and V, and  section VI concludes the paper.
\section{System Model}\label{sec1}
\subsection{Network and channel model}

In a distributed network, a receiver is surrounded by a number of
undesired source nodes that transmit on the same medium and in the
same time slot as its intended transmitter. Usually,  receiver
nodes do not  control  the number and the positions of these
sources of interference; rather, these  are determined by the
network dynamics and the access strategy in place. In the
particular case of a random access strategy, a simple way to model
the network is to consider interfering nodes distributed according
to a homogeneous Poisson point process (PPP) \cite{sous, ilow}. A
PPP relies on a single parameter $\lambda$, namely, the density of
transmitters. In a planar network,  it is described as follows:
 In  a closed region with area $A$, the number of transmitters
is distributed according to the Poisson law with density $\lambda
A$.  The positions of these nodes are uniformly distributed on the
plane. We assume that each node transmits with a single antenna
and receives with $L$ antennas. The distance between an emitting
node and its intended receiver is set to  $d_r$. Formally, the set
of emitting nodes forms a homogeneous PPP $\Pi=\{ X_i \in
\mathbb{R}^2, \lambda, i \in \mathbb{N} \}$, where $X_i$ are the
nodes' locations. The channel is modeled by two independent
 components, the first of which  represents the path-loss attenuation with an exponent $\alpha
> 2$, and  the second, represents the  channel coefficients, which
are independent among antennas and nodes.

\subsection{Interference expression under antenna array processing}

Since the network is modeled as a homogenous PPP, the interference
distribution does not depend on the spatial locations of
receivers. Thus, in the following we focus, without any loss of
generality, on a representative receiver placed at the origin. A
receiver is subject to a number of interfering signals coming from
non-desired transmitters. The received signal vector is then:

\begin{equation}
\mathbf{x}=d_r^{-\alpha/2}\mathbf{c}_r s_r+\sum_{X_k \in \Pi}
|X_k|^{-\alpha/2}\mathbf{c}_ks_k+\mathbf{n},
\end{equation}
where $\mathbf{c}_r$ and  $\mathbf{c}_k$, $k \in \mathbb{N}$, are
the propagation vectors with dimension $L$ that have independently
and identically distributed, zero-mean, unit variance  complex
Gaussian entries. The vector $\mathbf{n}$ is a zero-mean complex
Gaussian noise with variance $\sigma^2$ per dimension. All nodes
are supposed to use the same transmission power, normalized to
unity.
\\In statistical antenna array processing, a weight vector is
chosen based on the statistics of the data received and optimized
under a given criterion. Hence, the output of the antenna array
processor is:

\begin{equation}
\mathbf{y}=\mathbf{w}^T\mathbf{x},
\end{equation}
where $\mathbf{w}$ is a complex weight vector. The operator $~^T$
denotes the transpose conjugate. Several criteria could be
considered in determining this vector \cite{bair}, such as a
maximization of  the desired signal-to-noise ratio corresponding
to the maximal ratio combining receiver. In systems with
interference, the aim is to optimize the antenna array output such
that the quantity of noise and interference is minimized in the
resulting signal. In other words, maximizing the SINR:

\begin{equation}
\mathbf{w}_{OC}=\arg \max_{\mathbf{w}} \left\{\frac{\mathbf{w}^T
\mathbf{c}_r \mathbf{c}_r^T
\mathbf{w}}{\mathbf{w}^T(\mathbf{R}_I+\sigma^2\mathbf{I}_L)\mathbf{w}}\right\},
\end{equation}
where $\mathbf{I}_L$ is the $L \times L$ identity matrix and
$\mathbf{R}_I$ is the interference covariance matrix expressed as
$\mathbf{R}_I=\sum_{X_k \in \Pi}
|X_k|^{-\alpha}\mathbf{c}_k\mathbf{c}_k^T$. The well-known
solution  corresponds to \cite{bair,cox}:

\begin{equation}
\mathbf{w}_{OC}=(\mathbf{R}_I+\sigma^2\mathbf{I}_L)^{-1}\mathbf{c}_r.
\end{equation}
The resulting SINR is  denoted as $\beta$ and expressed as:
\begin{equation}
\beta=d_r^{-\alpha}\mathbf{c}_r^T\mathbf{R}^{-1}\mathbf{c}_r,
\end{equation}
where $\mathbf{R}$ is the interference plus noise covariance
matrix expressed as
$\mathbf{R}=\mathbf{R}_I+\sigma^2\mathbf{I}_L$.

\section{Outage probability derivation}\label{sec2}

The configuration we consider has two random parameters, namely,
the locations of interferers and the  channel coefficients. Thus,
we have a complex probabilistic system, which we propose to break
down into two levels. First, we generate the SINR expression
conditioned on a realization of the network, after which we
average over this random quantity. Consider an arbitrary planar
region $D$ with finite radius $d$ and  arbitrary nodes positions,
and use $N$ to denote the number of nodes in place. The problem
now is to find the distribution (or the eigenvalue distribution)
of random quadratic matrices having the form
$(\mathbf{C}\mathbf{P}\mathbf{C}^T+\sigma^2 \mathbf{I}_L)$. Each
column of  $\mathbf{C}$ represents a channel vector of an
interferer. Thus, the random matrix $\mathbf{C}$ has column
vectors independently and identically distributed as multivariate
normals with zero-mean and unit-variance vectors. The matrix
$\mathbf{P}=diag[{|X_1|^{-\alpha},|X_2|^{-\alpha}, \cdots }]$ is a
diagonal matrix  with real elements corresponding to the set of
received powers.
\subsection{Related results}
 The problem stated in the previous paragraph is a classic problem in the probability and wireless communication literature. We can
classify the most pertinent results to the context of this work
into two categories. The first category  concerns the work of
Silverstein and Bai \cite{bai}, who consider an
 asymptotic regime (the number of nodes and the number of antennas tend to infinity at a fixed rate). They established
that the eigenvalues of the considered class of random matrices
converge to a deterministic limiting distribution. This result is
first  used in \cite{tse1} and \cite{verdu2} to derive a closed
form expression of the asymptotic mean and variance of the SINR at
the output of the MMSE receiver, in the case of deterministic
received powers. In \cite{asym}, the authors apply these
asymptotic techniques  to the particular case of emitting nodes
uniformly distributed on an infinite plane,  and provide the SINR
mean and variance. Following this methodology, \cite{asym2} gives
an approximation of the SINR distribution based on the assumption
that the latter is a Gamma distribution. In our case, despite the
large number of users (the network area is wide), these results
cannot be applied. In fact, the spatial separations of  users
ensures that the global interference is only influenced by a small
number of them \cite{weber,olfa,serkey}. In addition, the goal of
our work is to
show how using a small number of receive antennas can improve the SINR.\\
The second category was initiated by Khatri  \cite{khatri}, who
derived the distribution of the matrix
$\mathbf{C}\mathbf{P}\mathbf{C}^T$. From this, \cite{shah}
provides the SIR distribution when  the noise is ignored:

\begin{equation}\label{khatri}
f_\beta(\beta)=\frac{\Gamma(N+1)}{\Gamma(L)\Gamma(N+1-L)}\frac{\beta^{L-1}q^{N+1}}{(1+q\beta)^{N+1}}|\mathbf{P}|^{-1}\mathrm{H}_0^{(N)}(N+1;
\mathbf{I}_N-q\mathbf{P}^{-1},\mathbf{Z}),
\end{equation}
where $H_0^{(N)}$ is a hypergeometric function of matrix
arguments, $\mathbf{Z}=diag[(1+q\beta)^{-1},\mathbf{I}_{L-1}]$,
$q$ is a particular constant and $\Gamma(\cdot)$ is the gamma
function.\\ Even if we consider that our system is
interference-limited,  expression (\ref{khatri}) is not simple as
it contains the hypergeometric function with matrix arguments. It
therefore appears to be difficult to derive  the SINR distribution
when $\mathbf{P}$ is a random matrix.   In \cite{gao1} and
\cite{gao2}, the authors extended the previous result by including
the noise term  and derived a  simpler expression. The derivation
consists in expressing the hypergeometric function as a polynomial
ratio. The  cumulative distribution function (CDF)  of the SINR is
given as:

\begin{equation}\label{equa:no2}
F_{\beta}(\beta | N, X_1 \cdots
X_N)=1-\frac{\sum_{i=0}^{L-1}a_i(\beta
d_r^{\alpha})^i}{\exp{(\sigma^2\beta
d_r^{\alpha})}\prod_{j=1}^N(1+|X_j|^{-\alpha}\beta d_r^{\alpha})},
\end{equation}
where $a_i$, $i=0\cdots L-1$, are the first $L$ coefficients of
the Taylor expansion of
$\exp(\sigma^2\beta)\prod_{j=1}^N(1+|X_j|^{-\alpha}\beta)$.
\subsection{Outage probability expression}
First, let us consider the case of a single receive antenna that
corresponds to  a Rayleigh fading channel. In this case, as
established by \cite{Baccelli}, the complementary CDF of the SINR
is  the characteristic function of the sum of interference and
noise, which is a perfect match for expression (\ref{equa:no2}).
In fact, $\prod_{j=1}^N(1+|X_j|^{-\alpha}\gamma)^{-1}$ is the
characteristic function expression of a weighted sum (with weights
equal to $|X_j|^{-\alpha}$) of $N$ independently and exponentially
distributed random variables.
 Since  the characteristic function of a Poisson class of interferers is well known \cite{schoky,rice,wil,lowen,syn,ilow,petro}, the
 derivation for $L=1$ is trivial. For an arbitrary number of antennas,
taking the expectation over the possible realizations of the
network and denoting $\beta d_r^{\alpha}$ by $\gamma$, we
establish the following theorem.
\begin{theorem}
Using the MMSE receiver, the outage probability in a Poisson field
of interferers and Rayleigh fading channel is:
\begin{eqnarray}\label{extreme}
F(\gamma,\lambda) =1- \sum_{i=0}^{L-1} \frac{(\lambda \Delta
\gamma^{2/\alpha}+\sigma^2\gamma)^i}{i!} \exp(-\lambda \Delta
\gamma^{2/\alpha}-\sigma^2\gamma),
\nonumber \\
\end{eqnarray}
where $\Delta=\pi 2/\alpha\Gamma(2/\alpha)\Gamma(1-2/\alpha)$.
\end{theorem}
\begin{IEEEproof}
 The coefficients $a_i$ in (\ref{equa:no2}) can be derived simply. In fact, extending the exponential in the denominator
of (\ref{equa:no2}) and putting $P_j=|X_j|^{-\alpha}$ we get:
\begin{equation}
\exp(\sigma^2\gamma)\prod_{j=1}^N(1+P_j\gamma)=\sum_{k=0}^{\infty}\frac{(\sigma^2)^k}{k!}\gamma^k
\sum_{i=0}^N b_i(P_1,\cdots,P_N) \gamma^i,
\end{equation}
where $b_i(P_1,\cdots,P_N)=\sum_{1\leq j_1<\cdots<j_i\leq
N}P_{j_1}P_{j_2}\cdots P_{j_i}$. Then, $a_i$ are expressed as
follows:
\begin{eqnarray}\label{equa:no3}
a_i=\sum_{k=0}^i \frac{(\sigma^2)^{i-k}}{(i-k)!}
b_k(P_1,\cdots,P_N) &&  i=1\cdots L-1,
\end{eqnarray}
From (\ref{equa:no2}) and  (\ref{equa:no3}):
\begin{eqnarray}\label{equa:no4}
F_\gamma(\gamma,\lambda) & = & 1-\exp(-\sigma^2\gamma)\mathrm{E}_N \left [ \sum_{i=0}^{L-1} \sum_{k=0}^{\min(i,N)} \frac{(\sigma^2)^{i-k}}{(i-k)!} \gamma^{i-k}\mathrm{E}_{P_1,\cdots, P_N}\left[ \frac{b_k(P_1,\cdots,P_N) \gamma^k}{\prod_{j=1}^N(1+P_j\gamma)}\right ]\right],\nonumber \\
\end{eqnarray}
where $\mathrm{E}_x$ denotes the expectation with respect to the
random variable $x$. The coefficients $b_k(P_1,\cdots,P_N)$,
$k=1\cdots L-1$, are composed of a sum of products of combinations
of $k$ elements from the set $\{P_j, j=1 \cdots N\}$.  Given that
the locations of the nodes are independently and identically
distributed, the last expression simplifies to yield:
\begin{eqnarray}\label{equa:no5}
\mathrm{E}_{P_1,\cdots, P_N}\left[
\frac{b_k(P_1,\cdots,P_N)\gamma^k}{\prod_{j=1}^N(1+P_j\gamma)}\right
] &=&C_N^k \mathrm{E}_{P_1,\cdots, P_N}\left[\frac{P_1 \cdots P_k
\gamma^k}{\prod_{j=1}^k(1+P_j
\gamma)}\frac{1}{\prod_{j=k+1}^N(1+P_j\gamma)}\right] \nonumber \\
&=& C_N^k \mathrm{E}_{P_1}\left[ \frac{P_1
\gamma}{(1+P_1\gamma)}\right]^k\mathrm{E}_{P_1}\left[
\frac{1}{(1+P_1\gamma)}\right]^{N-k}, \nonumber \\
\end{eqnarray}
where $C_N^k$ is the number of combinations of size $k$ from a set
of $N$ elements. To compute $\mathrm{E}_{P_1}\left[
\frac{P_1}{(1+P_1\gamma)}\right]$, recall that conditioned on N,
in a finite region $D \in \mathbb{R}^2$, the locations of nodes
are uniformly distributed. Consequently:
\begin{equation}\label{eq131}
\mathrm{E}_{P_1}\left[
\frac{P_1}{(1+P_1\gamma)}\right]=\frac{1}{\pi
d^2}\int_{D}|X|^{-\alpha}(1+|X|^{-\alpha}\gamma)^{-1}\mathrm{d}X.
\end{equation}
Similarly, we have:
\begin{equation}\label{eq141}
\mathrm{E}_{P_1}\left[ \frac{1}{(1+P_1\gamma)}\right]=\frac{1}{\pi
d^2}\int_{D}(1+|X|^{-\alpha}\gamma)^{-1}\mathrm{d}X.
\end{equation}
Since the number of nodes is distributed according to the Poisson
law with mean $\lambda \pi d^2$, and considering (\ref{eq131}) and
(\ref{eq141}), the CDF of the SINR yields:
\begin{eqnarray}
F_\gamma(\gamma,\lambda) & = &1-
\exp(-\sigma^2\gamma)\sum_{N=0}^\infty \sum_{i=0}^{L-1}
\sum_{k=0}^{\min(i,N)} \frac{N!}{k!(N-k)! (i-k)!} (\sigma^2)^{i-k}
\gamma^{i-k} \left( \frac{1}{\pi d^2} \int_D
\frac{|X|^{-\alpha}\gamma}{1+|X|^{-\alpha}\gamma}\mathrm{d}X\right)^k  \cdot \nonumber \\
                 & & \left(
\frac{1}{\pi d^2} \int_D
\frac{\mathrm{d}X}{1+|X|^{-\alpha}\gamma}\right)^{N-k} \frac{(\lambda \pi d^2)^N}{N!}\exp(-\lambda \pi d^2)\nonumber \\
&=&1-\exp(-\sigma^2\gamma)\sum_{i=0}^{L-1}
\sum_{k=0}^i\sum_{N=k}^\infty \frac{(\sigma^2 \gamma)^{i-k}}{k!
(i-k)!}\left( \lambda \int_D
\frac{|X|^{-\alpha}\gamma}{1+|X|^{-\alpha}\gamma}\mathrm{d}x\right)^k \cdot \nonumber \\
& & \frac{1}{(N-k)!}\left( \lambda \int_D
\frac{\mathrm{d}X}{1+|X|^{-\alpha}\gamma}\right)^{N-k}\exp(-\lambda
\pi d^2).
\end{eqnarray}
Using the fact that $\pi d^2=\int_D 1 \mathrm{d}X$, we get: \small
\begin{equation}\label{equa:no6}
 F_\gamma(\gamma,\lambda)  = 1-\exp(-\sigma^2\gamma)\sum_{i=0}^{L-1} \sum_{k=0}^i
\frac{1}{k! (i-k)!} (\sigma^2 \gamma)^{i-k}  \left( \lambda \int_D
\frac{|X|^{-\alpha}\gamma}{1+|X|^{-\alpha}\gamma}\mathrm{d}X\right)^k
\exp( \lambda \int_D
(\frac{1}{1+|X|^{-\alpha}\gamma}-1)\mathrm{d}X ).
\end{equation}
\normalsize
 Finally, letting the area of $D$ go to infinity and evaluating   the integrals
in (\ref{equa:no6}) (the complete evaluation of these integrals is
provided in the Appendix) we get:
\begin{eqnarray}\label{equa:no22}
F_\gamma(\gamma,\lambda)
&=&1-\exp(-\sigma^2\gamma)\sum_{i=0}^{L-1} \sum_{k=0}^i
\frac{1}{k! (i-k)!} (\sigma^2 \gamma)^{i-k} \left(
\gamma^{2/\alpha} \lambda \Delta\right)^k \exp( - \gamma^{2/\alpha}\lambda \Delta )\nonumber \\
&=&1- \sum_{i=0}^{L-1}  \frac{(\lambda \Delta \gamma^{2/\alpha}
 + \sigma^2 \gamma )^i }{i!}  \exp( -  \lambda \Delta \gamma^{2/\alpha}-\sigma^2 \gamma ). \nonumber \\
\end{eqnarray}
\end{IEEEproof}

\section{Discussion and application }
\subsection{Noise-limited regime and interference-limited regime}
 The expression  (\ref{extreme}) derived for
the outage probability has a simple form. From it, we can easily
see the effects of various parameters on the network performance.
Moreover, it clearly presents  the trade-off between noise and
interference cancellation, and thus provides some insight into the
intuition behind the result.

\subsubsection{Noise-limited regime}
In  systems where the density of users  is negligible, the SNR
cumulative distribution function becomes:
\begin{equation}\label{mrc}
F_\gamma(\gamma)=1-\sum_{i=0}^{L-1} \frac{(\sigma^2\gamma)^i}{i!}
\exp(-\sigma^2\gamma).
\end{equation}
As expected, (\ref{mrc}) is the classic expression of a Chi-square
cumulative distribution function. The latter provides the SNR
distribution when maximal ratio combining is performed, that is
equivalent to optimum combining in a complex Gaussian noise
environment.
\subsubsection{Interference-limited regime} Where the noise is
negligible, the CDF is:
\begin{equation}\label{int}
F_\gamma(\gamma,\lambda)=1-\sum_{i=0}^{L-1} \frac{(\lambda \Delta
\gamma^{2/\alpha})^i}{i!} \exp(- \lambda \Delta
\gamma^{2/\alpha}).
\end{equation}
Equation (\ref{int}) corresponds, up to the factor $\Delta$, to
the probability that the $L^{th}$ largest received power is below
the threshold $\gamma$ \cite{david}. This is not surprising given
some properties of the interference. Since the path loss exponent
is greater than $2$,  the  interference distribution is heavy
tailed \cite{murd}. Therefore,  there is a large dispersion
between the   received powers. Let us order the received powers
and denote them as $|X_{(1)}|^{-\alpha}\gg |X_{(2)}|^{-\alpha}\gg
|X_{(3)}|^{-\alpha}\gg \cdots$. With probability one, the random
matrix $\mathbf{C}\mathbf{P}\mathbf{C}^T$ is of order $L$, and its
eigenvalues could be approximated as $|X_{(i)}|^{-\alpha}$,
$i=1\cdots L$ \cite{eric}. Thus the maximal eigenvalue of the
inverse of  $\mathbf{C}\mathbf{P}\mathbf{C}^T$ is approximately
$|X_{(L)}|^{\alpha}$.  The performance is then primarily affected
by the $L^{th}$ strongest received power, and so, the outage is
almost due to the $L^{th}$ strongest interferer, and corresponds
to the event $|X_{(L)}|< \gamma^{1/\alpha}$. In other words,  to
achieve successful reception, the $L^{th}$ strongest interferer
must be outside the region of radius $\gamma^{1/\alpha}$. This
fact matches expression (\ref{int}) up to the scalar factor
$\Delta$.
\subsubsection{SIR mean and variance}
From (\ref{int}), we can establish in a straightforward manner
that in an interference-limited regime, the mean and the variance
of the SIR are, respectively:
\begin{equation}\label{mean}
E[SIR]=\frac{\Gamma(L+\alpha/2)}{(L-1)!}\frac{d_r^{-\alpha}}{\lambda^{\alpha/2}\Delta^{\alpha/2}},
\end{equation}
\begin{equation}
Var[SIR]=\left(\frac{\Gamma(L+\alpha)}{(L-1)!}-\left[\frac{\Gamma(L+\alpha/2)}{(L-1)!}\right]^2\right)\frac{d_r^{-2\alpha}}{\lambda^{\alpha}\Delta^{\alpha}}.
\end{equation}
Thus, the MMSE receiver provides an antenna array gain equal to
$\Gamma(L+\alpha/2)/(L-1)!$. It should be noted that expression
(\ref{int})  is consistent with the result previously established
in \cite{asym}. In fact, when the number of antennas is
sufficiently high, the mean SIR is approximately equal to
$L^{\alpha/2}\frac{d_r^{-\alpha}}{\lambda^{\alpha/2}\Delta^{\alpha/2}}$.
This latter relation is also derived for the asymptotic regime,
i.e., very large $L$, in \cite{asym}.
%\subsubsection{Transmission capacity under outage constraint}
%The transmission capacity under outage constraint is defined in
%\cite{weber5} as the maximum density of transmitters  ensuring
%that the outage probability does not exceed a fixed constant
%$\epsilon$. Considering a small outage constraint and using
%(\ref{extreme}) we have:
%\begin{corollary}
%The transmission capacity under a small outage constraint
%$\epsilon$ satisfies:
%\begin{equation}
%\lambda_{\epsilon}\approx \frac{(L!
%\epsilon)^{1/L}}{c(\alpha)\beta^{2/\alpha}d_r^2}
%\end{equation}
%where $c(\alpha)=\pi 2/\alpha \Gamma(2/\alpha)\Gamma(1-2\alpha)$.
%\end{corollary}
%\subsubsection{Single hop throughput capacity}
\subsection{Application: single-hop throughput capacity}
The interference model considered serves to capture several
network classes under some additional assumptions.  In the
following, we focus  on a single-hop ad hoc network with an ALOHA
 access protocol. Thus, the density of interferers represents the spatial rate at which
transmissions occur, i.e., the contention density. In this
context, the outage probability represents the spatial average of
the density of communications that fail to be established at a
given range $d_r$ in almost every given realization of the network
\cite{Baccelli}. Equally, the mean number of successful
transmissions or the throughput per unit area is
$T=\lambda(1-F(\gamma,\lambda))$. The simplicity of our analytical
result  allows the direct optimization of the contention density
as:
\begin{equation}
\lambda_{max}=\arg \max_{0\leq \lambda< \infty}
{\lambda(1-F(\gamma,\lambda))}.
\end{equation}
\begin{corollary}\label{cor}
In an interference-limited regime, the optimum contention density
is:
\begin{equation}
\lambda_{max}=\frac{g(L)}{\Delta \gamma^{2/\alpha}}.
\end{equation}
The parameter $g(L)$ is only a function of the number of receive
antennas and corresponds to the positive root of the following
polynomial:
\begin{equation}\label{poly}
Q(t)=\sum_{i=0}^{L-1} \frac{t^i}{i!}-\frac{t^L}{(L-1)!}
\end{equation}
Moreover, this parameter satisfies:
\begin{equation}
\frac{L}{2} \leq g(L) \leq L
\end{equation}
The equivalent spatial throughput per unit area is:
\begin{equation}
T_{max}=\frac{g^{L+1}(L)}{(L-1)!\Delta\gamma^{2/\alpha}}\exp(-g(L)).
\end{equation}
\end{corollary}
Corollary \ref{cor} indicates that the MMSE receiver provides a
linear scaling of the optimum contention density with the number
of receive antennas unlike to the MRC and the zero-forcing
receivers, whose scaling laws are $L^{2/\alpha}$ and
$L^{1-2/\alpha}$, respectively \cite{weber4,weber3}. The linear
scaling law achieved by the MMSE receiver is predicted in
\cite{weber6}. In fact, the authors show that  linear scaling is
possible  with a partial zero-forcing receiver, which is
suboptimal as compared to the MMSE receiver. Through corollary
\ref{cor}, we confirm this prediction, and we show that the
scaling law is exactly linear.

%For examples, with one antenna $g(1)$ is equal to $1$ and with two
%antennas this parameter becomes $g(2)=\frac{1+\sqrt{5}}{2}$.
%The last expression  serves to quantify  the potential improvement
%of the spatial throughput by the use of antenna array processing.
%This improvement is equals to
%$\frac{g^{L+1}(L)}{(L-1)!}\exp(-g(L)+1)$. For example, with only
%two antennas the spatial throughout is increased by a factor of
%$2.28$.

\section{Simulation Results}
\begin{figure}[!t]
\centering
\includegraphics[width=5.65in, height=5.1in]{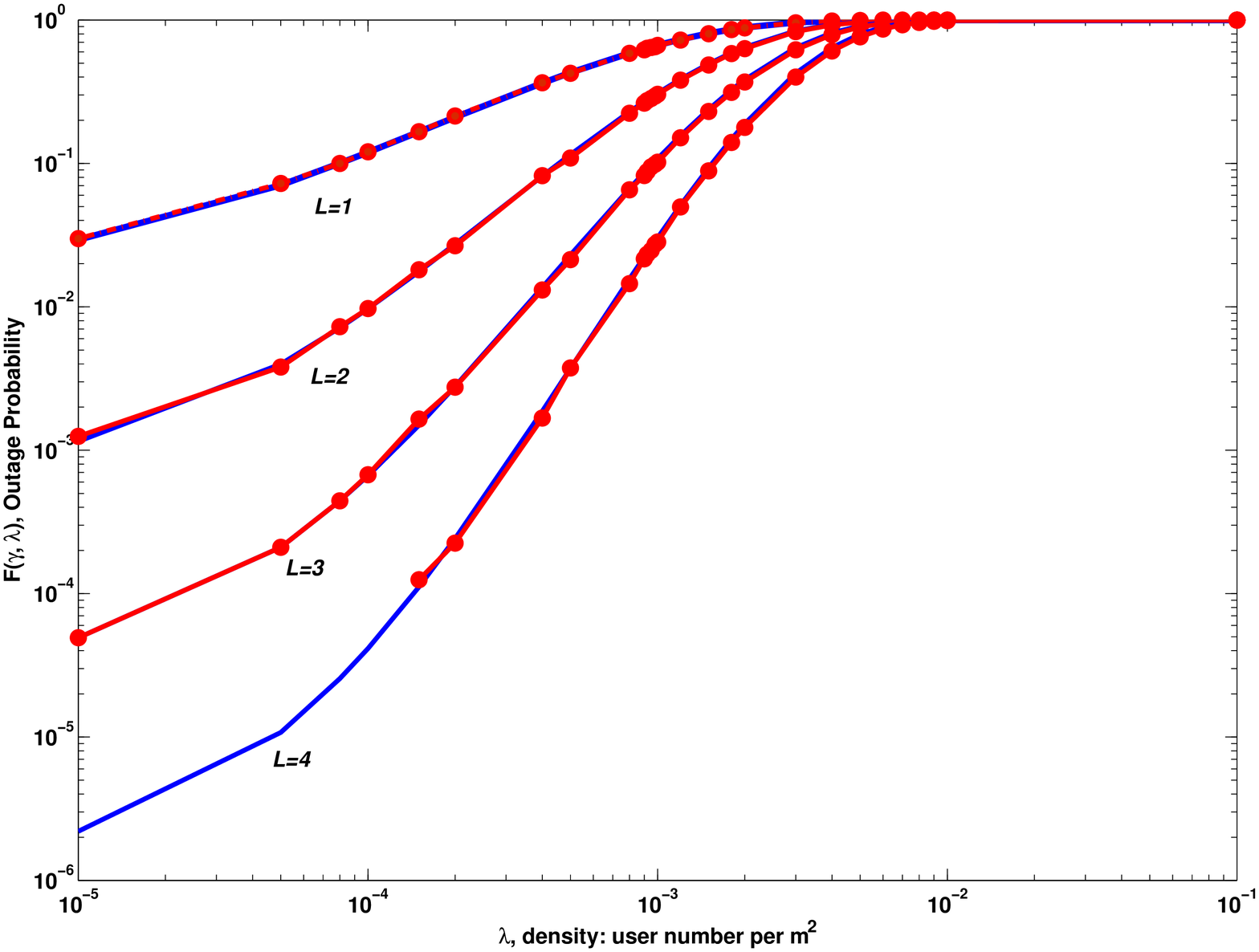}
\caption{Outage probability: simulation (dashed lines) and
theoretical (solid lines) results with $\sigma^2=-50 dB/antenna$
(all nodes use unit transmission power.)} \label{figsim1}
\end{figure}

\begin{figure}[!t]
\centering
\includegraphics[width=5.65in, height=5.1in]{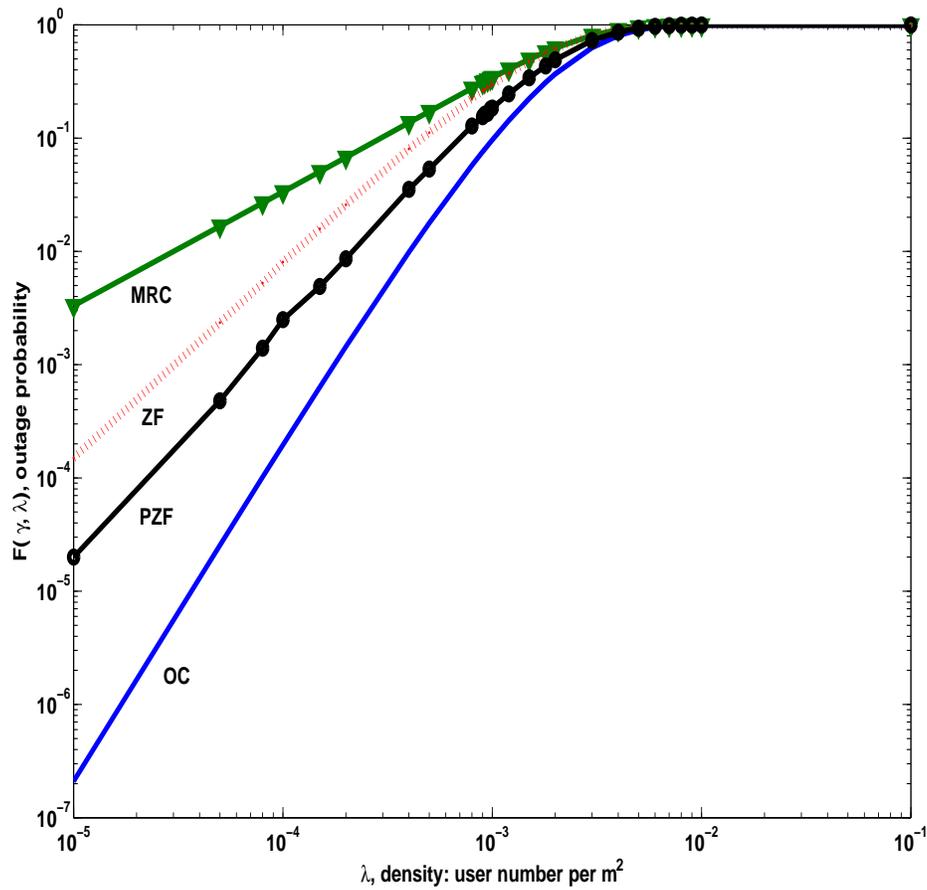}
\caption{Outage probability: Maximal ratio combining, zero
forcing, partial zero forcing (with the optimum number of
cancelled interferers established in \cite{weber6}) and optimum
combining
 with $L$=3 and $\sigma^2=0$.} \label{figsim3}
\end{figure}

\begin{figure}[http]
\centering
\includegraphics[width=5.65in, height=5.1in]{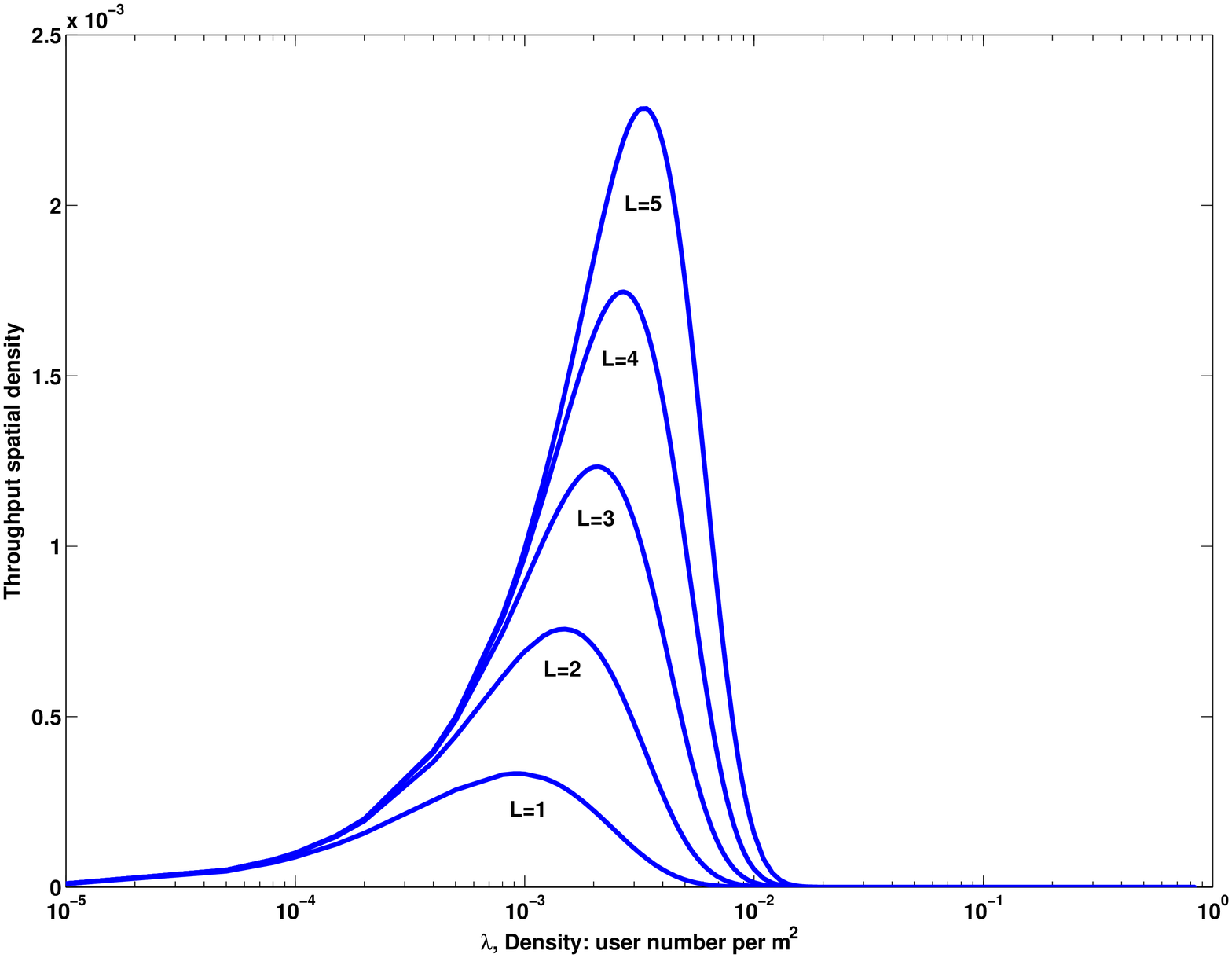}
\caption{Optimum combining: Throughput spatial density with
$L=1\cdots 5$, $\sigma^2=-57dB/antenna$.} \label{figsim4}
\end{figure}

\begin{figure}[!t]
\centering
\includegraphics[width=5.65in, height=5.1in]{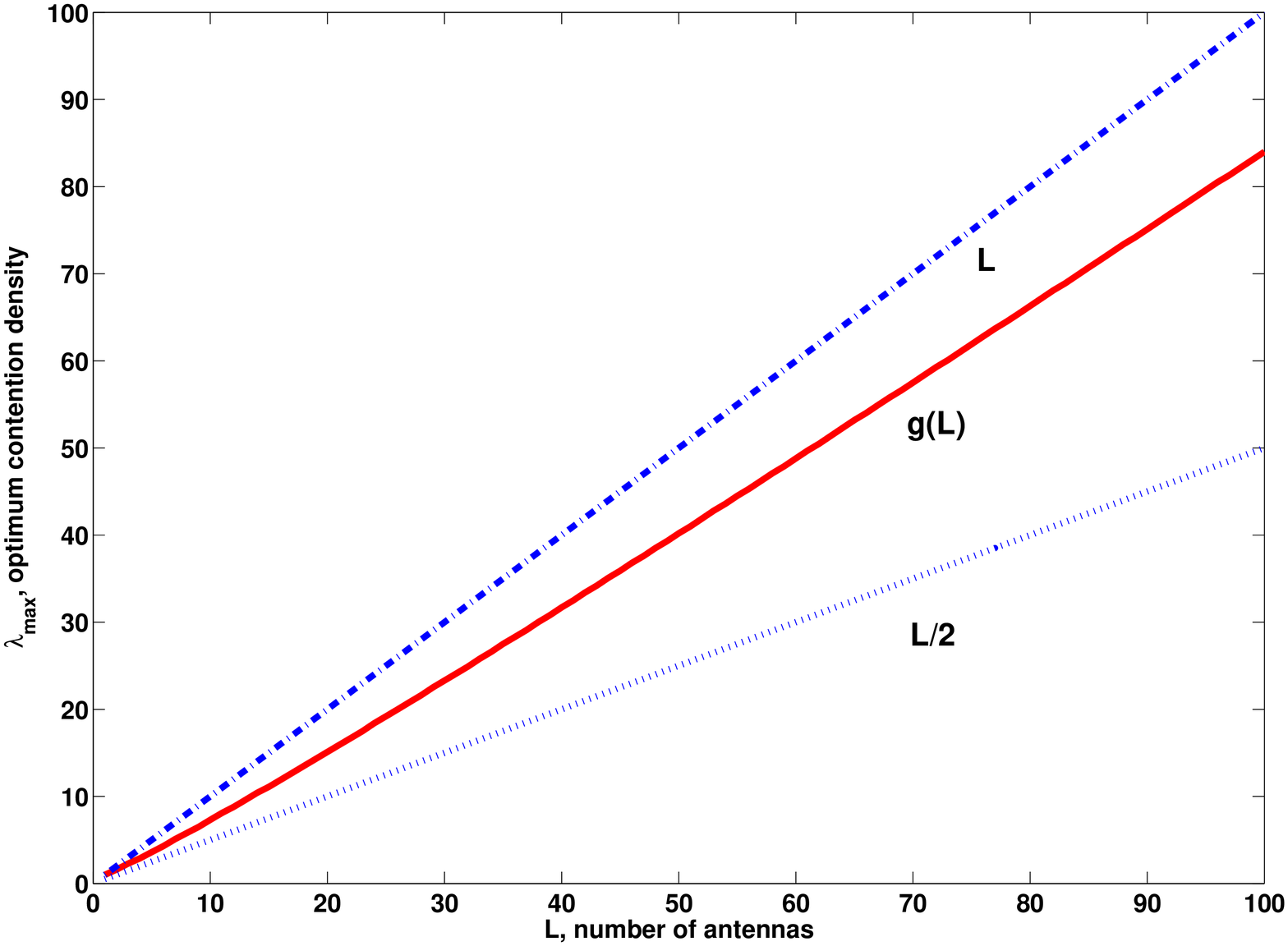}
\caption{Optimum combining: Optimum contention density as a
function of the antenna array dimension, with $\sigma^2=0$ and
$\Delta \gamma^{2/\alpha}$ normalized to $1$.} \label{figsim5}
\end{figure}

All   simulations are carried out with the following parameters:
path-loss coefficient $\alpha=3.5$, SINR threshold $\beta=3 dB$
and distance between transmitter and receiver $d_r=10 m$. Figure
\ref{figsim1}  shows the simulation and the analytical results of
the outage probability as a function of the density $\lambda$
where $1,2,3$ and $4$ antennas  are used. The Monte Carlo
simulation and the analytical curves are very close, with the gap
between them  arising  from the fact that  the analysis performed
on the previous section concerns infinite networks. In our
simulations, the area  is finite, and depends on the density
$\lambda$. It  is chosen such that we have $100$ emitters in the
area,  on  average. In \cite{ilow2},  the author provides a
complete analysis of the error on the interference estimation when
a finite network is considered rather than  an infinite one, and
establishes that this error depends on the exponent $\alpha$.\\
Figure \ref{figsim3}  provides a performance comparison between
the optimal combining receiver and three antenna array processing
techniques, namely, maximal ratio combining, zero-forcing and
partial zero-forcing receivers. It is clear that the optimum
combining receiver outperforms the  other techniques. This is due
to the fact that maximal ratio combining deals only with the
fading effects of the desired signal. Zero-forcing receiver uses
all additional degrees of freedom provided by the antennas to
cancel  strong interferers, while the partial zero-forcing
receiver uses some of the antennas for  interference cancellation
and provides diversity with the remaining ones; they are
nevertheless still suboptimal as compared to the MMSE receiver,
which provides the best trade-off between interference
cancellation and spatial diversity. \\
 Figure \ref{figsim4} presents the throughput density
improvement with the number of antennas and figure \ref{figsim5}
shows the linear scaling of the optimum contention density.

\section{Conclusion}
This paper derived the exact outage probability of the optimum
combining receiver in the presence of noise and  Poisson field of
interferers.  The framework developed is nonspecific and can be
generalized in a straightforward manner to any
  fading having a power distribution of the form
$\sum_k x^k e^{-kx}$, such as the Nakagami fading. The result
obtained has a simple closed form, and provides an understanding
of   the performance of the MMSE receiver, in addition to allowing
a comparison  with other antenna array processing methods, such as
maximal ratio combining and zero-forcing receivers. The assumption
made on the interference is pertinent to many network classes. In
addition, the result can be used to study the performance of the
optimum combining receiver at the access and at the network
layers. Finally, the simulations provided in the paper demonstrate
that experimental and theoretical results match perfectly.

\ifCLASSOPTIONcaptionsoff
  \newpage
\fi

\appendices
\section{Evaluation of the integrals in
(\ref{equa:no6})}
%\subsection{Evaluation of the integrals in (\ref{equa:no6})}
 The integrals in (\ref{equa:no6}) are evaluated as follows:
\begin{eqnarray}\label{equa:no15}
\int_{\mathbb{R}^2} ((1+|X|^{-\alpha}\gamma)^{-1} -1)\mathrm{d}X & = &  \int_0^\infty \int_0^{2\pi}((1+r^{-\alpha}\gamma)^{-1} -1)r \mathrm{d}\theta \mathrm{d}r \nonumber \\
                   & = &  -2\pi\int_0^\infty \frac{r^{-\alpha+1}}{1+r^{-\alpha}\gamma} \mathrm{d}r. \nonumber \\
\end{eqnarray}

Putting $y=r^{\alpha}\gamma$, the last expression becomes:

\begin{eqnarray}\label{equa:no16}
\int_{\mathbb{R}^2} ((1+|X|^{-\alpha}\gamma)^{-1} -1)\mathrm{d}X & = &  -2\pi\frac{\gamma^{2/\alpha}}{\alpha}\int_0^\infty \frac{y^{1-2/\alpha}}{1+y} \mathrm{d}y \nonumber \\
                                                       & = &  -2\pi\frac{\gamma^{2/\alpha}}{\alpha}\Gamma(2/\alpha)\Gamma(1-2/\alpha) \nonumber \\
                                                       & = &  -\gamma^{2/\alpha}\Delta. \nonumber \\
\end{eqnarray}

On the other hand, we have
\begin{equation}
\int_{\mathbb{R}^2}
\frac{|X|^{-\alpha}\gamma}{1+|X|^{-\alpha}\gamma}\mathrm{d}x=-\int_{\mathbb{R}^2}
((1+|X|^{-\alpha}\gamma)^{-1} -1)\mathrm{d}x,
\end{equation}
\section{Proof of corollary 1}
By Descartes' rule of signs, the polynomial in expression
(\ref{poly}) has at most one real positive root. The value of this
polynomial at $L$ is negatively signed:
\begin{eqnarray}
Q(L)&=&\sum_{i=0}^{L-1} \frac{L^i}{i!}-\frac{L^L}{(L-1)!} \nonumber \\
&=& \sum_{i=0}^{L-1}\left[ \frac{L^i}{i!}-\frac{L^{L}}{L!}\right]
\nonumber \\
&=& \sum_{i=0}^{L-1}\left[
\frac{L^i}{i!}\left(1-\frac{L^{L-i}}{(i+1)\cdots L}\right) \right]
\nonumber \\
&\leq& 0
\end{eqnarray}
On the other hand, the value of the polynomial at $L/2$ is lower
bounded  as follows:
\begin{eqnarray}
Q(L/2)&=&\sum_{i=0}^{L-1} \frac{L^i}{2^i i!}-\frac{L^L}{2^L(L-1)!} \nonumber \\
&\geq& \sum_{i=0}^{L-1} C_L^i \frac{1}{2^i}-\frac{L^L}{2^L(L-1)!} \nonumber \\
&\geq&
\left(\frac{3}{2}\right)^L-\frac{1}{2^L}-\frac{L^L}{2^L(L-1)!}\label{in}
\end{eqnarray}
The right hand side of inequality (\ref{in}) is positive for all
values of $L$.  Thus, the considered polynomial has one positive
root that lies in the interval $[L/2,L]$.
\section*{Acknowledgment}

The authors would like to thank Dr. Sergey Loyka for his helpful
comments and suggestions.
\bibliographystyle{IEEEtran}
\bibliography{IEEEabrv,articlejroptimumv2}

\newpage

~~~\\

~~~\\
~~~\\
~~~\\
%\begin{figure*}[ht]
%\centerline{\subfloat[ $\sigma^2=-5.7 dB$]{
%\includegraphics[width=4.15in, height=3.5in]{resultf1.eps}
%\subfloat[ $\sigma^2=-5 dB$]{\includegraphics[width=4.15in, height=3.5in]{resultf2.eps}%
%\label{fig_second_case}}} \caption{Outage probability: simulation
%(dashed lines) and theoretical (solid lines) results}
%\label{figsim1}
%\end{figure*}

\newpage
~~~
\\

~~~
\\
~~~
\\
~~~
\\

~~~
\\

\newpage

%\begin{biography}[{\includegraphics[width=1in,height=1.25in,clip,keepaspectratio]{mshell}}]{Michael Shell}

%\end{IEEEbiography}


%\newpage



%\vfill


\end{document}